\documentstyle[12pt]{article} 
\def\ro#1{\hbox{\rm #1}}
\def\is#1{{\it #1\/}}
\def\bo#1{{\bf #1}}

\font\teneuf=eufm10 scaled \magstep1 \font\seveneuf=eufm7 scaled \magstep1
\font\fiveeuf=eufm5 scaled \magstep1
\newfam\euffam
\textfont\euffam=\teneuf \scriptfont\euffam=\seveneuf
   \scriptscriptfont\euffam=\fiveeuf

\def\frak#1{{\fam\euffam\relax#1}}

\def\fg{\frak g} \def\fh{\frak h}

\def\Cal#1{{\cal#1}}

\def\CD{\Cal D} \def\CF{\Cal F} \def\CH{\Cal H} 
\def\CN{\Cal N} \def\CP{\Cal P} 
 \def\CU{\Cal U}

\def\barr#1{\overline{#1}{}}
\def\sbar{\bar s{}}\def\Vbar{\,\overline{\!V}{}}
\def\xbar{\bar x{}}
 \def\Ghat{\widehat G{}}
\def\Phat{\widehat P{}}

\def\comp{\raise 1pt \hbox{$\,\scriptstyle\circ\,$}}

\def\hexnumber#1{\ifcase#1 0\or1\or2\or3\or4\or5\or6\or7\or8\or9\or
 A\or B\or C\or D\or E\or F\fi}

\font\tenmsb=msbm10 scaled \magstep1\font\sevenmsb=msbm7 scaled \magstep1
\font\fivemsb=msbm5 scaled \magstep1
\newfam\msbfam
\textfont\msbfam=\tenmsb  \scriptfont\msbfam=\sevenmsb
  \scriptscriptfont\msbfam=\fivemsb

\edef\msbhx{\hexnumber\msbfam}

\def\Bbb#1{{\fam\msbfam\relax#1}}

\def\R{\Bbb R}
\def\C{\Bbb C}
\def\N{\Bbb N}
\def\K{\Bbb K}

\def\Id{\hbox{\tenrm 1\kern-3.8pt \elevenrm1}}

\def\hexnumber#1{\ifcase#1 0\or1\or2\or3\or4\or5\or6\or7\or8\or9\or
 A\or B\or C\or D\or E\or F\fi}

\font\eightex=cmex10 scaled 800 \font\sixex=cmex10 scaled 600
  \font\fourex=cmex10 scaled 400
\newfam\sexfam
\textfont\sexfam=\eightex  \scriptfont\sexfam=\sixex
  \scriptscriptfont\sexfam=\fourex
\edef\sexhx{\hexnumber\sexfam}

\mathchardef\odotsy="220C
\mathchardef\oplussy="2208
\mathchardef\bigvee="1\sexhx57
\mathchardef\bigwedge="1\sexhx56
\mathchardef\bigcap="1\sexhx54
\mathchardef\bigcup="1\sexhx53
\mathchardef\bigotimes="1\sexhx4E
\mathchardef\bigoplus="1\sexhx4C
\mathchardef\bigodot="1\sexhx4A

\def\odot{\raise 1pt \hbox{$\,\scriptscriptstyle\odotsy\,$}}
\def\tensor{\raise 1pt \hbox{$\,\scriptscriptstyle\otimes\,$}}
\def\Wedge{\raise 1pt \hbox{$\bigwedge$}}
\def\Odot{\raise 1pt \hbox{$\bigodot$}}
\def\Oplus{\raise 1pt \hbox{$\bigoplus$}}
\def\Tensor{\raise 1pt \hbox{$\bigotimes$}}

\def\comp{\raise 1pt \hbox{$\,\scriptstyle\circ\,$}}

\mathchardef\semidirect="2\msbhx6E
\mathchardef\directsemi="2\msbhx6F
\mathchardef\emptyset="0\msbhx3F

\def\GL#1{\ro{GL}(#1)} \def\gl#1{\frak{gl}(#1)}
\def\GLR#1{\GL{#1,\R}} \def\glR#1{\gl{#1,\R}}
 
  \def\sL#1{\frak{sl}(#1)}

  \def\so#1{\frak{so}(#1)}

\let\nn\nonumber

\def\roq#1{\quad \ro{#1}\quad }
\def\roqq#1{\qquad \ro{#1}\qquad }

\def\ie{i.e., }
\def\eg{e.g., }
\def\cf{cf.~}

\def\f#1{{\textstyle {1\over #1}}}
\def\fr#1#2{{\textstyle {#1\over #2}}}

\def\rg#1#2{#1=1,\ldots,#2}

\def\sups #1#2{#1^1,\ldots,#1^{#2}}
\def\psups #1#2{(#1^1,\ldots,#1^{#2})}

\def\aeq#1{\ba #1\ea}

\def\ba{\begin{eqnarray}}
\def\ea{\end{eqnarray}}

\def\set#1#2{\mathchoice{\left \{#1\vphantom{#2}\;\right |
\left .\;#2\vphantom{#1}\right \}}{\{#1\vphantom{#2}\,|\,#2\vphantom{#1} \}}
{\{#1\vphantom{#2}\,|\,#2\vphantom{#1} \}}
{\{#1\vphantom{#2}\,|\,#2\vphantom{#1} \}}}

\def\operator#1{\expandafter\def\csname#1\endcsname{\mathop{\ro{#1}}\nolimits}}
\operator{Ker}\operator{Span}\operator{Div}
\operator{sech} \operator{csch}
\operator{arcsinh} \operator{arccosh}
\operator{sn}  \operator{cn}  \operator{dn}
\operator{diag}

\def\odo#1{\mathchoice{d \over d#1}{d/d#1}{d/d#1}{d/d#1}}

\def\odow#1{\mathchoice{d^2 \over d#1^2}%
{d^2 /d#1^2}{d^2 /d#1^2}{d^2 /d#1^2}}
\def\pd#1#2{\mathchoice{\partial #1 \over \partial #2}%
{\partial #1/\partial #2}{\partial #1/\partial #2}{\partial #1/\partial #2}}
\def\pdo#1{\mathchoice{\partial \over \partial #1}%
{\partial /\partial #1}{\partial /\partial #1}{\partial /\partial #1}}
\def\pdow#1#2{\mathchoice{\partial^2 \over \partial #1 \partial #2}%
{\partial^2 /\partial #1 \partial #2}{\partial^2 /\partial #1 \partial #2}%
{\partial^2 /\partial #1 \partial #2}}

\def\px{\partial _x} \def\qy{\partial _y}

\def\ip#1#2{\langle #1\,; #2\rangle}

\def\bov{\bo v}
\def\J#1{\ro J{}^{#1}}

\newcount\itemnumber \itemnumber=1
\def\ctitle#1.{\begin{center}\large
#1\end{center}\par\nobreak\vskip-10pt\nobreak}
\def\btitle#1.{{\center{\bf #1}\par\nobreak}}

\def\ntable#1{\itemnumber=1 
   \def\\{\cr \the\itemnumber.\global\advance\itemnumber by 1
&}\vcenter{\openup1\jot 
   \halign{##&&\quad $\displaystyle{}##$\hfill\cr 
   \the\itemnumber.\global\advance\itemnumber by 1 &#1\cr}}}

\def\Lntable#1{\begin{enumerate}%
\def\\{$\item\quad$\displaystyle} \def\p{\hfill\break\null\kern20pt}
\item\quad$#1$\end{enumerate}}

\newcounter{mylc}
\def\items#1{\begin{list}{\themylc .}%
   {\usecounter{mylc}\settowidth{\labelwidth}{iiii)}}%
   \def\\{\item}\\#1\end{list}}

\def\Section#1#2.{\section{#2}\label{Sc:#1}\def\sectionkey{#1}}
\def\sc#1{Section~\ref{Sc:#1}}

\def\Eq#1{\label{eq:\sectionkey.#1}}

\def\first#1.#2\ends{#1}
\def\second#1.#2\ends{#2}
\def\eq#1{(\ref{eq:\process{#1.@\ends}})}
\def\process#1{\if @\second #1\sectionkey.\first #1\else
                    \first #1.\expandafter\first\second #1\ends\fi}

\def\be{\begin{equation}}
\def\ee{\end{equation}}

\newtheorem{thm}{\bf Theorem}
\newtheorem{lemma}[thm]{\bf Lemma}

\newenvironment{defn}{\par\medskip\refstepcounter{thm}\noindent{\bf Definition
\thethm}\enspace}{\par\medskip\noindent}
\newenvironment{rem}{\par\medskip\noindent{\it
Remark:}\enspace}{\par\medskip\noindent}

\def\Th#1 #2\par{\begin{thm}\label{Th:\sectionkey.#1}#2\end{thm}}
\def\Lm#1 #2\par{\begin{lemma}\label{Lm:\sectionkey.#1}#2\end{lemma}}
\def\Df#1 #2\par{\begin{defn}\label{Df:\sectionkey.#1}#2\end{defn}}
\def\Remark #1\par{\begin{rem}#1\end{rem}}

\def\th#1{Theorem~\ref{Th:\process{#1.@\ends}}}

\def\df#1{Definition~\ref{Df:\process{#1.@\ends}}}

\def\nosemiz#1;{#1}
\def\key#1 #2\par{\bibitem{#1}#2\par}

\newif\ifnames \namesfalse
\def\namez#1;{\namesfalse\namezz#1,;} 
\def\namezz#1,#2,#3;{\ifx,#3,\ifnames \ and\fi\fi #1,#2,\ifx,#3,\else 
    \namestrue\namezz #3;\fi}

\def\book#1;#2; #3\par{\namez#1;{\em #2} (#3).}
\def\paper#1;#2;#3; #4 (#5)#6--#7\par{\namez#1;{\em #3}\if a#4, to appear.\else
    { \bf #4},#6 (#5).\fi}
\def\preprint#1;#2;#3\par{\namez#1;#2, preprint,#3.}
\def\inbook#1;#2;#3; #4\par{\namez#1; {\sl in:}{\em #3} (#4).}
\def\other#1\par{#1.}

\renewcommand{\thefootnote}{\dag}

\let\la\lambda
\newcommand{\PSI}{\Phi}
\def\dx{\partial _x} \let\Dx\dx \def\dy{\partial _y} \def\Dz{\partial_z}

\def\qes.{quasi-exactly solvable}
\def\sch.{Schr\"odinger}
\def\dfo{differential operator}
\def\JJ#1{\,(J^{#1})^2}
\def\p#1{\CP_{#1}}
\let\De\Delta \let\al\alpha  
\def\Ah{\hat A}

\def\Jp#1{J^+_{#1}}
\def\J0#1{J^0_{#1}}
\def\Jm{J^-}
\def\Te{T^\epsilon}
\def\Tp{T^+}
\def\T0{T^0}
\def\Tm{T^-}
\def\Qa{Q_\alpha}

\def\bpm{\left(\begin{array}{cc}} \def\epm{\end{array}\right)}
\def\Ut{\tilde U}
\def\Wt{\tilde W}
\def\pb{\bar p}
\def\Qb{\,\overline{\!Q}{}}
\def\Qab{\,\overline{\!Q}{}_{\al}}
\def\qb#1#2#3{\,\overline{\!q}{}_{#1}(#2,#3)}
\def\qab#1#2{\qb{\al}{#1}{#2}}

\begin{document}
\bibliographystyle{unsrt}
\setcounter{page}1

\pagestyle{myheadings}
\markboth{\hfill {Finkel, Gonz\'alez-L\'opez, Kamran, Olver and
Rodr\'\i guez}}{\uppercase{Lie
Algebras and Partial Integrability} \hfill}
\thispagestyle{empty}

\vbox{\vspace{6mm}}
\begin{center}
\renewcommand{\thefootnote}{\fnsymbol{footnote}}
{\large \bf\uppercase{Lie Algebras of Differential
Operators}\\[2mm]\uppercase{and Partial Integrability}%
\footnotemark[4]%
\footnotetext[4]{Talk presented at the IV Workshop on Differential Geometry and
its Applications (Santiago de Compostela, Spain, 18--20 September
1995).}}\\[1cm]
Federico Finkel\footnote{Departamento de F\'\i sica Te\'orica
II, Universidad Complutense, 28040 Madrid, SPAIN. Supported in part by DGICYT
grant no.~PB92--0197.}\\[2pt]
Artemio Gonz\'alez-L\'opez\footnotemark[1]\\[2pt] Niky
Kamran\footnote{Department of Mathematics, McGill University, Montr\'eal,
Qu\'ebec, CANADA H3A 2K6. Supported in part by a NSERC Grant.}\\[2pt]
Peter J.~Olver\footnote{School of Mathematics, University of
Minnesota, Minneapolis, MN, USA 55455. Supported in part by NSF grants DMS
92--04192 and 95--00931.}\\[2pt]
Miguel A. Rodr\'\i guez\footnotemark[1]
\end{center}
\vspace{2mm}

\begin{abstract}
This paper surveys recent work on
Lie  algebras of differential operators and their application to
the construction of quasi-exactly solvable Schr\"odinger operators.
\end{abstract}


\Section{intro}Introduction.
Lie-algebraic and Lie group theoretic methods have played a significant role
in the  development of quantum mechanics since its inception.  In the
classical applications, the Lie group  appears as a symmetry group of the
Hamiltonian operator, and the associated representation theory  provides an
algebraic means for computing the spectrum.  Of particular importance are the
exactly  solvable problems, such as the harmonic oscillator or the hydrogen
atom, whose point spectrum can  be completely determined using purely
algebraic methods.  The fundamental concept of a ``spectrum  generating
algebra'' was introduced by Arima and Iachello, \cite{ArIai}, \cite{ArIaii},
to study nuclear  physics problems, and subsequently, by Iachello, Alhassid,
G\"ursey, Levine, Wu and their collaborators, was  also successfully applied
to molecular dynamics and spectroscopy, \cite{IaLe},
\cite{Levine}, and  scattering theory, \cite{AEW}, \cite{AGIa}, \cite{AGIb}. 
The Schr\"odinger operators amenable to the  algebraic approach assume a
\is{Lie-algebraic form,} meaning that they belong to the universal  enveloping
algebra of the spectrum generating algebra.  Lie-algebraic operators
reappeared in the  discovery of Turbiner, Shifman, Ushveridze, and their
collaborators, \cite{Shifman}, \cite{ShifTurb}, \cite{Turbiner}, \cite{Ush},
of a new class of physically significant spectral problems, which they  named
\is{quasi-exactly solvable,} having the property that a (finite) subset of the
point spectrum can be  determined using purely algebraic methods.  This is an
immediate consequence of the additional  requirement that the hidden symmetry
algebra preserve a finite-dimensional representation space  consisting of
smooth wave functions.  In this case, the Hamiltonian restricts to a linear
transformation  on the representation space, and hence the associated
eigenvalues can be computed by purely  algebraic methods. 
Connections with conformal field theory, \cite{Gorsky},  \cite{MPRST},
\cite{Shifcft}, \cite{zabro}, and the theory of orthogonal polynomials,
\cite{Turbpoly},
\cite{Turbpolz}, \cite{Tuone}, lend additional impetus for the
study of such problems.

In higher dimensions, much less is known than in the one-dimensional case; in
fact, only a few special examples of quasi-exactly solvable problems in two
dimensions have appeared in the literature to  date, \cite{ShifTurb},
\cite{GKOii}.
Complete lists of  finite-dimensional Lie algebras of differential operators
are known in two (complex) dimensions;  there are essentially 24 different
classes, some depending on parameters.  The quasi-exactly solvable  condition
imposes a remarkable quantization constraint on the cohomology parameters
classifying  these Lie algebras.  This phenomenon of the ``quantization of
cohomology'' has recently been  given an algebro-geometric interpretation,
\cite{GHKO}.  Any of the resulting  quasi-exactly solvable  Lie algebras of
differential operators can be used to construct new examples of
two-dimensional  quasi-exactly solvable spectral problems. An additional
complication is that, in higher dimensions,  not every elliptic second-order
differential operator is equivalent to a Schr\"odinger operator (\ie  minus
Laplacian plus potential), so not every Lie-algebraic operator can be assigned
an immediate  physical meaning.  The resulting ``closure conditions'' are
quite complicated to solve, and so the  problem of completely classifying
quasi-exactly solvable Schr\"odinger operators in two dimensions  appears to
be too difficult to solve in full generality.  A variety of interesting
examples are given in 
\cite{GKOii}, and we present a few particular cases of interest here.

The above ideas, originally introduced for scalar Hamiltonians describing
spinless particles, can be generalized to include spin.
The first step in this direction was taken by Shifman and Turbiner,
\cite{ShifTurb}, using the fact that a Hamiltonian for a spin $1/2$
particle in $d$ spatial dimensions can be constructed from a Lie
superalgebra of first order \dfo s in
$d$ ordinary (commuting) variables and one Grassmann (anticommuting)
variable. Alternatively, \cite{BrKo}, $2 \times 2$ matrices (or
$N\times N$ matrices for particles of arbitrary spin, \cite{BGGK}) can
be used to represent the Grassmann variable. In  contrast
with the scalar case, very few examples of matrix \qes. \sch. operators
have been found thus far, \cite{ShifTurb}. There are important conceptual
reasons for this fact. First of all, in the matrix case Lie
superalgebras of matrix differential operators come naturally into play,
while in the scalar case only Lie algebras need be considered. Secondly, as we
shall see in
\sc l, every scalar second order
\dfo{} in one dimension can be transformed into a
\sch. operator of the form
$-\partial_x^2+V(x)$ by a suitable change of the independent variable $x$
and a local rescaling of the wave function. For matrix \dfo s, on the other
hand, the analogue of this result --- $V(x)$ being now a Hermitian matrix
of smooth functions --- is no longer true unless the operator satisfies
quite stringent conditions.

In the last section of this paper, we will discuss
the characterization of the class of \qes. matrix differential operators
preserving a finite-dimensional space of wave functions with polynomial
components (\cf \cite{Tumat}, \cite{BrKo}, \cite{BGGK}).
For the important particular case of spin
$1/2$ particles, we will give necessary and
sufficient conditions for a \qes. operator to be equivalent to a non-trivial
\sch. operator. A suitable simplification of these conditions will then be
used to construct new examples of multi-parameter \qes. spin $1/2$
Hamiltonians in one dimension.

\Section{qes}Quasi-Exactly Solvable Schr\"odinger Operators. Let $M$ denote an
open subset of Euclidean space $\R^d$ with coordinates $x = \psups xd$.  The 
time-independent Schr\"odinger equation for a differential operator $\CH$ is
the eigenvalue problem
\be \CH\cdot\psi = \lambda \psi. \Eq7\ee  In the quantum mechanical
interpretation, $\CH$ is a (self-adjoint) second-order differential
operator, which plays the role of  the quantum ``Hamiltonian'' of the system. 
A nonzero wave function
$\psi (x)$ is called {em  normalizable} if it is square integrable, \ie lies
in the Hilbert space $\ro L^2(\R^d)$, and so represents  a physical bound
state of the quantum mechanical system, the corresponding eigenvalue
determining  the associated energy level. Complete explicit lists of
eigenvalues and eigenfunctions are known for only a  handful of classical
``exactly solvable'' operators, such as the harmonic oscillator.  For the
vast  majority of quantum mechanical problems, the spectrum can, at best, only
be approximated by  numerical computation. The quasi-exactly solvable systems
occupy an important intermediate station, in that a finite part of the
spectrum can be computed by purely algebraic means.

To describe the general form of a quasi-exactly solvable operator, we begin
with a finite-dimensional Lie algebra $\fg$ spanned by $r$ linearly
independent first-order differential operators
\be
J^a = \sum_{i=1}^d \xi^{ai}(x) \pdo{x^i} + \eta^a(x), \quad\rg ar,
\Eq{20}
\ee whose coefficients $\xi^{ai}, \eta^a$ are smooth functions of $x$. Note
that each differential operator  is a sum, $J^a = \bo v^a + \eta ^a$, of a
{\it vector field} $\bo v^a = \sum \xi^{ai}
\pdo{x^i}$  (which may be zero) and a {\it multiplication operator} $\eta ^a$.

A differential operator is said to be {\it Lie-algebraic} if it lies in the
universal enveloping  algebra $\CU(\fg)$ of the Lie algebra $\fg$, meaning
that it can be expressed as a polynomial in the  operators $J^a$.  In
particular, a  second-order differential operator $T$ is Lie-algebraic if it
can be  written as a quadratic combination
\be -T = \sum _{a, b} c_{ab} J^a J^b + \sum _a c_a J^a + c_0, \Eq9\ee
for
certain constants $c_{ab}$, $c_a$, $c_0$.  (The minus sign in front of the
Hamiltonian is taken  for later convenience.) Note that the commutator
$[J^c,T]$ of the Hamiltonian with any generator of $\fg$, while  still of the
same Lie-algebraic form, is not required to vanish.  Therefore, the {\it
hidden symmetry algebra} $\fg$ is not a  symmetry algebra in the traditional
sense.  Lie-algebraic operators appeared in the early work of  Iachello,
Levine, Alhassid, G\"ursey and collaborators in the algebraic approach to
scattering theory,  \cite{AEW},
\cite{AGIa}, \cite{AGIb}, \cite{Levine}. The condition of quasi-exact
solvability imposes an additional constraint on the Lie algebra and hence  on
the type of operators which are allowed.  A Lie algebra of first order
differential operators $\fg$  will be called {\it quasi-exactly solvable} if
it possesses a finite-dimensional representation space (or  module) $\CN
\subset C^\infty$ consisting of smooth functions; this means that if $\psi \in
\CN$ and 
$J^a \in \fg$, then $J^a(\psi) \in \CN$.  A differential operator $T$ is
called {\it quasi-exactly  solvable} if it lies in the universal enveloping
algebra of a quasi-exactly solvable Lie algebra of  differential operators. 
Clearly, the module  $\CN$ is an invariant space for $T$, 
\ie $T(\CN)\subset \CN$, and hence $T$ restricts to a linear matrix operator
on $\CN$.  We will  call the eigenvalues and corresponding eigenfunctions for
the restriction $T | \CN$ {\it algebraic,}  since they can be computed by
algebraic methods for matrix eigenvalue problems.  (This does not  mean that
these functions are necessarily algebraic in the traditional pure mathematical
sense.)   Note that the number of such algebraic eigenvalues and
eigenfunctions equals the dimension of 
$\CN$. So far we have not imposed any normalizability conditions on the
algebraic eigenfunctions,  but, if they are normalizable, then the
corresponding algebraic eigenvalues give part of the point  spectrum of the
differential operator.

It is of great interest to know when a given differential operator is in Lie
algebraic or quasi-exactly  solvable form.  There is not, as far as we know,
any direct test on the operator in question that will answer this in general. 
Consequently, the best approach to this problem is to effect a complete 
classification of such operators under an appropriate notion of equivalence. 
In order to classify Lie  algebras of differential operators, and hence Lie
algebraic and quasi-exactly solvable Schr\"odinger  operators, we need to
precisely specify the allowable equivalence transformations.  
\Df1 Two differential operators $T(x)$ and $\barr T(\xbar)$ are {\it
equivalent\/} if they can be mapped into each other by a  combination of
change of independent variable,
\be \xbar = \varphi (x), \Eq1\ee
and {\it gauge transformation}
\be \barr T(\xbar) = e^{\sigma (x)}\cdot  T(x)\cdot e^{-\sigma (x) }.\Eq2\ee

The transformation \eq1--\eq2 have two key properties.  First, they respect
the commutator between  differential operators, and therefore preserve their
Lie algebra structure. In particular, if $\fg$ is a finite-dimensional Lie
algebra of first-order differential operators then the transformed algebra
$\barr\fg = \set{\barr T}{T\in\fg}$ is a Lie algebra isomorphic to
$\fg$. Moreover, if
$\CN$ is a finite-dimensional $\fg$-module then $\barr\CN=\set{e^{\sigma(x)}
f(x)|_{x=\varphi^{-1}(\xbar)}}{f\in\CN}$ is a finite-dimensional
$\barr\fg$-module. In other words, if $\fg$ is quasi-exactly solvable so is
$\barr\fg$.  It immediately follows that the transformation \eq1--\eq2
preserves the class of \qes. operators: in other words, if $T(x)$ is
\qes. with respect to $\fg$ then the transformed operator $\barr T(\xbar)$
will be \qes. with respect to the Lie algebra $\barr\fg\simeq\fg$.

Second, \eq1--\eq2 preserve the  spectral problem \eq7 associated to the
differential operator $T$, so that if
$\psi (x)$ is an  eigenfunction of $T$ with eigenvalue $\lambda $, then the
transformed (or ``gauged'') function
\be \barr \psi(\xbar) = e^{\sigma  (x)} \psi (x), \roqq{where} \xbar = \varphi
(x), \Eq3\ee is the corresponding eigenfunction of $\barr T$ having the same
eigenvalue. Therefore, this notion  of equivalence is completely adapted to
the problem of classifying quasi-exactly solvable  Schr\"odinger operators.
The gauge factor $\mu (x) = e^{\sigma (x)}$ in \eq2 is {\it not} necessarily 
unimodular, \ie $\sigma  (x)$ is not restricted to be purely imaginary, and
hence does not necessarily  preserve the normalizability properties of the
associated eigenfunctions.  Therefore, the problem of  normalizability of the
resulting algebraic wave functions must be addressed.

Let us summarize the basic steps that are required in order to obtain a complete classification of
quasi-exactly solvable operators and their algebraic eigenfunctions:

\items{Classify finite-dimensional Lie algebras of differential operators.\\
Determine which Lie algebras are quasi-exactly solvable.\\
Solve the equivalence problem for differential operators.\\
Determine normalizability conditions.\\
Solve the associated matrix eigenvalue problem.}

In this survey we will concentrate on the solution of the first three
problems, referring the reader to \cite{GKOnorm} for the solution of the
normalizability problem in the one-dimensional case.

\Section{edo}Equivalence of Differential Operators.
Consider a second-order linear differential operator
\be  -T = \sum_{i,j=1}^d g^{ij}(x)\pdow{x^i}{x^j} + \sum_{i=1}^d h^i(x) \pdo{x^i} + k(x), 
\Eq1\ee
defined on an open subset $M\subset \R^d$.  We are interested in studying the problem of when two 
such operators are equivalent under the combination of change of variables and gauge
transformation
\eq{qes.1}, \eq{qes.2}.  Of particular importance is the question of when $T$ is equivalent to a 
{\it Schr\"odinger operator,\/} which we take to mean an operator  $\CH = -\Delta +V(x)$, where
$\Delta$  denotes either the flat space Laplacian or, more generally, the Laplace-Beltrami
operator over a curved  manifold. This  definition of Schr\"odinger operator
excludes the introduction of a magnetic field, which, however,  can also be handled
by these methods. As we shall see, there is an essential difference
between one-dimensional and  higher dimensional spaces in the solution to the
equivalence problem for second-order differential  operators, because in higher
space dimensions not every second-order differential  operator is locally equivalent
to a Schr\"odinger operator $-\Delta +V(x)$, where $\Delta$ is the flat space
Laplacian.

Explicit equivalence conditions were
first found by \'E.~Cotton, \cite{Cotton}, in 1900.  Since the symbol of a linear differential
operator is invariant under coordinate  transformations, we begin by assuming that the operator is
elliptic, meaning that the symmetric matrix 
$\Ghat(x) = \bigl(g^{ij}(x)\bigr)$ determined by the leading coefficients of $-T$ is
positive-definite.  Owing to the induced transformation rules under the change of variables \eq1,
we can interpret  the inverse matrix $G(x) = \Ghat(x)^{-1}  = \bigl(g_{ij}(x)\bigr)$ as
defining a Riemannian  metric  
\be ds^2 = \sum _{i,j=1}^d g_{ij}(x) dx^i dx^j, \Eq3\ee 
on the subset $M\subset \R^d$.  We will follow the usual tensor convention of raising and lowering 
indices with respect to the Riemannian metric \eq3.  We rewrite the differential operator \eq1 in a 
more natural coordinate-independent form as
\be T = -\sum_{i,j=1}^n g^{ij} (\nabla_i-A_i)(\nabla_j-A_j) + V, \Eq4 \ee 
where $\nabla_i$ denotes covariant differentiation using the Levi-Civita connection associated
to the metric $ds^2$.   The vector $\bo A(x) = (A^1(x),\ldots ,A^d(x))$ can be thought of as a
(generalized) magnetic vector  potential; in view of its transformation properties, we define the
associated {\it magnetic one-form} 
\be  A = \sum_{i=1}^d A_i(x) \,dx^i. \Eq6\ee 
(Actually, to qualify as a physical vector potential, $\bo A$ must be purely imaginary and satisfy
the  stationary Maxwell equations, but we need not impose this additional physical constraint in
our  definition of the mathematical magnetic one-form \eq6.)  The explicit formulas relating the
covariant  form \eq4 to the standard form \eq1 of the differential operator can be
found in \cite{GKOii}.  Each second-order elliptic operator then is uniquely 
specified by a metric, a magnetic one-form, and a potential function $V(x)$. In
particular, if the  magnetic form vanishes, so $A=0$, then $T$ has the form of a
Schr\"odinger operator
$T =  -\Delta + V$, where $\Delta $ is the Laplace-Beltrami operator associated with
the metric \eq3.

The application of a gauge transformation \eq{qes.2} does not affect the
metric or the potential;  however, the magnetic one-form is modified by an exact one-form: $A
\mapsto  A  +  d\sigma  $.  Consequently, the  {\em magnetic two-form} $\Omega
=dA
$, whose coefficients  represent the associated magnetic field, {\it is}
unaffected by gauge transformations.

\Th2 Two elliptic second-order differential operators $T$ and $\barr T$ are (locally)
equivalent  under a change of variables $\xbar = \varphi (x)$ and gauge transformation \eq{qes.2}
if and only if  their metrics, their magnetic two-forms, and their potentials are mapped to each
other
\be\varphi^* \bigl(d\sbar^2\bigr) = ds^2, \qquad \varphi^* \bigl(\,\barr \Omega \,\bigr) = \Omega,
\qquad\varphi^* \bigl(\,\Vbar \,\bigr) = V.\Eq8\ee
{\em(}Here $\varphi^* $ denotes the standard pull-back action of $\varphi $ on
differential forms; in  particular, $\varphi^* (\Vbar) = \Vbar \comp 
\varphi$.{\em)}
\endgraf
In particular, an elliptic second-order differential operator is equivalent to a Schr\"o\-din\-ger
operator $-\Delta + V$ if and only if its magnetic one-form is closed: $dA  = \Omega =0$. 
Moreover, since the curvature tensor associated with the metric is invariant, 
$T$ will be equivalent to a ``flat" Schr\"odinger operator if and only if
the metric $ds^2$ is flat, \ie has  vanishing Riemannian curvature tensor,
and the magnetic one-form is exact.

In the one-dimensional case, every metric is automatically flat and all
1-forms are exact. Hence the previous theorem implies that every elliptic second-order
differential operator is equivalent to a flat Schr\"odinger operator. See \cite{GKOqes}
for explicit formulas for the change of variables, the gauge factor and
the potential.

\Section{lado}Lie Algebras of Differential Operators.
In this section, we summarize what is known about the classification problem for Lie algebras of
first  order differential operators.  Any finite-dimensional Lie algebra  $\fg$  of first order
differential  operators in $\K ^d$ (with $\K =\R$ or $\K =\C$) has a basis of the form
\ba &J^1 = \bo v^1 + \eta ^1(x), \ldots , J^r = \bo v^r + \eta ^r(x),&\nonumber\\
&J^{r+1} = f ^1(x), \ldots , J^{r+s} = f ^s(x), \Eq1 \ea
\cf \eq{qes.20}.  Here $\sups {\bo v}r$  are linearly independent vector fields spanning an 
$r$-dimensional Lie algebra  $\fh$.  The functions  $f ^1(x),\ldots , f ^s(x)$  define
multiplication  operators, and span an abelian subalgebra $\CF$ of the full Lie algebra $\fg$. 
Since the commutator 
$[\bov^i,f ^j] = \bov^i(f ^j)$ is a multiplication operator, which must belong to $\fg$,
we  conclude that $\fh$ acts on $\CF$, which is a finite-dimensional $\fh$-module
(representation  space) of smooth functions.  The functions $\eta ^a(x)$ must
satisfy additional constraints in order  that the operators \eq1 span a Lie
algebra; we find
\be [\bov^i + \eta ^i,\bov^j + \eta ^j] = [\bov^i ,\bov^j] + \bov^i (\eta ^j) - \bov^j (\eta ^i).
\Eq{01}\ee 
Now, since $\fh$ is a Lie algebra, $[\bov^i ,\bov^j] = \sum_k c_k^{ij} \bov^k$, where $c_k^{ij}$
are the  structure constants of $\fh$.  Thus the above commutator will belong to $\fg$ if and only
if
\be\bov^i (\eta ^j) - \bov^j(\eta^i) - \sum_k c_k^{ij} \eta ^k\in\CF.\ee
These conditions can be conveniently
re-expressed using the basic theory of Lie algebra cohomology, \cite{Jacobson}.  Define the {\it
one-cochain} $F$ on the Lie algebra of vector fields $\fh$ as the linear map  
$F\colon \fh\to  C^\infty\equiv C^\infty(\K ^d)$  which satisfies $\ip F{\bo v^a} = \eta ^a$.  Since
we can add in any  constant coefficient linear combination of the  $f ^b$'s  to the  $\eta ^a$'s 
without changing the  Lie algebra  $\fg$,  we should interpret the  $\eta ^a$'s  as lying in the
quotient space  $C^\infty / \CF$,  and hence regard  $F$  as a  $C^\infty / \CF$--valued cochain. 
In view of \eq{01}, the collection of differential  operators \eq1 spans a Lie algebra if and only
if the cochain $F$ satisfies 
\be \bo v \ip F{\bo w}  -  \bo w \ip F{\bo v}  -  \ip F{[\bo v, \bo w]} \in \CF, \qquad\forall\bo
v, \bo  w  \in   \fh.\Eq5\ee 
The left hand side of \eq5 is just the evaluation $\ip{\delta _1F}{\bo v, \bo w}$ of the {\it 
coboundary} of the 1-cochain  $F$, hence \eq5 expresses the fact that the cochain  $F$  must be a 
$C^\infty / \CF$--valued {\it cocycle}, \ie $\delta_1F=0$.  A 1-cocycle is itself a coboundary,
written
$F =
\delta _0\sigma  $ for some $\sigma (x)\in C^\infty $, if and only if $\ip F {\bo v} = \bo
v(\sigma )$ for all 
$\bo v\in \fh$, where $\bov(\sigma )$
is considered as an element of $C^\infty / \CF$.  It can be shown that two cocycles will differ by
a coboundary  $\delta _0\sigma$ if and only if the corresponding Lie algebras are equivalent
under the gauge transformation \eq{qes.2}.  Therefore two cocycles lying in the same {\it
cohomology class} in the cohomology space  $H^1(\fh, C^\infty / \CF ) = \Ker \delta_1 /
\ro{Im}\,\delta _0$  will give rise to equivalent Lie algebras of differential operators.  In 
summary, then, we have the following fundamental characterization of Lie algebras of first order 
differential operators:
\Th 1  There is a one-to-one correspondence between equivalence classes of
finite dimensional Lie  algebras $\fg$ of first order differential operators on  $M$  and
equivalence classes of triples  $\bigl(  \fh, \CF , [F] \bigr)$,  where
\items{$\fh$  is a finite-dimensional Lie algebra of vector fields,\\
$\CF  \subset  C^\infty$  is a finite-dimensional  $\fh$-module of functions,\\ $[F]$  is a
cohomology class in  $H^1(\fh, C^\infty / \CF )$.} \par Based on \th1, there are three basic steps
required to classify finite dimensional Lie algebras of first  order differential operators. 
First, one needs to classify the finite dimensional Lie algebras of vector  fields  $\fh$ up to
changes of variables; this was done by Lie in one and two complex dimensions, \cite{Liegr},
\cite{Lietrans}, and by the authors in two real dimensions, \cite{GKOreal}, under the regularity
assumption that the Lie algebra has no singularities --- not all vector fields in the Lie algebra 
vanish at a common point. Secondly, for each
of these  Lie algebras, one needs to classify all possible finite-dimensional  $\fh$-modules $\CF$ 
of  
$C^\infty$ functions. Finally, for each of the
modules  $\CF$, one needs to  determine the first cohomology space $H^1(\fh,
C^\infty / \CF )$.  As the tables at the end of \cite{GKOlado} indicate, the 
cohomology classes are parametrized by one or more continuous parameters or,
in a few cases, smooth functions.

It is then a fairly straightforward matter to determine when a given Lie algebra satisfies the
quasi-exact solvability condition that it admit a non-zero finite-dimensional module $\CN\subset
C^\infty $. Remarkably, in all known cases, the cohomology parameters are
``quantized'', the quasi-exact  solvability requirement forcing them to assume
at most a discrete set of distinct values.  This  intriguing phenomenon of
``quantization of cohomology'' has been geometrically explained in the
maximal cases in terms  of line bundles on complex surfaces in
\cite{GHKO}.

In one dimension, there is essentially only one quasi-exactly solvable Lie
algebra of first order differential operators:
\Th2 Every {\em(}non-singular{\em)} finite-dimensional quasi-exactly solvable
Lie algebra of first order  differential operators in one {\em(}real or
complex{\em)} variable is locally equivalent to a subalgebra of one of  the Lie
algebras
\be \widehat\fg_n   =    \Span \bigl\{\; \partial _x , \quad  x\partial _x ,\quad   x^2 \partial
_x  -  n\, x  ,\quad   1 \;\bigr\}\simeq\gl2, \Eq2\ee
where  $n\in \N$.  For $\widehat\fg_n $, the associated module $\CN = \CP_n$ consists of the 
polynomials of degree at most $n$.

Turning to the two-dimensional classification, a number of additional complications present 
themselves.  First, as originally shown by Lie, there are many more equivalence classes of
finite-dimensional Lie algebras of vector fields, of arbitrarily large dimension and sometimes
depending on arbitrary functions as well.  Moreover, the classification results in
$\R^2$ and $\C^2$  are no longer the same; in fact, the real classification has been completed
only very recently, \cite{GKOrqes}.  Another complication is that the modules $\CF$ for the
vector field Lie algebras are no longer necessarily spanned by monomials, a fact
that makes the  determination of the cohomology considerably more difficult.
Our classification results for
finite-dimensional Lie algebras of differential operators in two real or complex
variables, \cite{GKOCanada}, \cite{GKOlado}, \cite{GKOrqes}, are summarized in the Tables
appearing, respectively, in
\cite{GKOrqes} and \cite{GKOqes}.
The classification naturally  distinguishes
between the \is{imprimitive} Lie algebras, for which there exists an invariant 
foliation of the manifold, and the \is{primitive} algebras, having no such
foliation. Each of the Lie algebras appearing in the complex classification
has an obvious real counterpart, obtained by restricting the  coordinates to be
real.  Moreover, as explained in
\cite{GKOrqes}, in such cases the associated  real Lie algebras of
differential operators and finite-dimensional  modules are readily  obtained by
restriction.  As a matter of fact, every imprimitive real Lie algebra of
vector fields in $\R^2$ is obtained by this  simple procedure.  In addition,
there are precisely five primitive real Lie algebras of vector fields in two
dimensions that are are not equivalent under a {\it real} change of
coordinates to any of the Lie algebras obtained by straightforward restriction
of the complex normal forms. The complete list of these additional real forms
along with their canonical complexifications appear in Table 4 of
\cite{GKOrqes}. The maximal (complex) algebras,  namely Case 1.4 of
\cite{GKOrqes} ($\sL2\oplus \sL2$), Case 2.3 ($\sL3$), and Case 1.11
($\gl2\semidirect \R^r$) play an  important role in Turbiner's theory of
differential equations in two dimensions with orthogonal  polynomial
solutions, \cite{Turbopc}. Of the additional real normal forms, Case 4.2
($\so3$) was extensively used by Shifman and Turbiner, \cite{ShifTurb}, while
Case 4.5 ($\so{3,1}$) is related to recent examples of quasi-exactly solvable
Hamiltonians in two dimensions constructed by Zaslavskii, \cite{zasso31},
\cite{zastwod}.

\Section{l} Quasi-Exactly Solvable Operators on the Line.
Let us now specialize to problems in one
dimension.  In view of \th{lado.2}, we let $n\in 
\N$ be a nonnegative integer, and consider the Lie algebra $\fg_n$ spanned by
the  differential operators
\be J^- =J^-_n = \odo z ,\qquad J^0 = J^0_n =z \odo z  -{n\over 2}, \qquad J^+ =  J^+_n= z^2 \odo
z   -  n z ,\Eq3\ee
which satisfy the standard $\sL2$ commutation relations.  In this section, we
shall use $z$  instead of $x$ for the ``canonical coordinate'', retaining $x$ for the physical
coordinate in  which the operator takes Schr\"odinger form. Since $\fg_n$ differs from the Lie
algebra $\widehat \fg_n$ in \th{lado.2} only by the inclusion of constant
functions, any Lie-algebraic operator 
\eq{qes.9} for the full algebra $\widehat \fg_n$ is automatically a Lie-algebraic operator for 
the subalgebra $\fg_n$. Therefore, in our analysis of Lie-algebraic differential operators we can,
without loss of generality, concentrate on the Lie algebra $\fg_n$.

Using \eq3, it is readily seen that the most general second-order quasi-exactly
solvable operator in one space dimension can be written in the canonical form
\be-T   =  P \odow z  +  \left \{Q - {n-1\over 2}P'\right \} \odo z 
+\left
\{R  -{n\over 2}Q' + {n(n-1)\over 12}P''\right \},\Eq1\ee
where $P(z)$, $Q(z)$, and $R(z)$ are (general) polynomials of respective
degrees
$4, 2, 0$, whose explicit expression in terms of the constants $c_{ab}$ and
$c_a$ can be found in
\cite{GKOqes}.  Since the module
$\CN$ is the  space
$\CP_n$ of polynomials of degree at most $n$, the algebraic eigenfunctions of \eq1  will, in the
$z$-coordinates, just be polynomials $\chi _k(z)\in \CP_n$.  In terms of the  standard basis
$\nu _k(z) = z^k$, $k = 0,\ldots ,n$, the restriction $T \mid 
\CP_n$ takes the form of a pentadiagonal matrix. In summary, for a normalizable one-dimensional
quasi-exactly solvable  operator there are $n+1$ algebraic eigenfunctions which, in the canonical
$z$ coordinates, are polynomials of degree at most $n$.

Specializing the solution to the equivalence problem given by \th{edo.2} to the operator
\eq1, we find (\cf \cite{GKOqes}) that if $P(z)>0$ the change of variables
required to place the operator into physical (Schr\"odinger) form will, in general, be
given by an elliptic integral
\be x  =  \varphi (z)  =  \int^z {dy \over \sqrt{P(y)}} \;, \Eq{10}\ee
the corresponding gauge factor being
\be\mu (z)  = P(z)^{-n/4}\, \exp \left\{ \int^z {Q(y)\over 2P(y)} \, dy \right\} . 
\Eq{11}\ee
The potential is given by
\be V(x)  =  - \frac{n(n+2)\left (P P'' - \fr34P'^2\right ) + 3(n+1) ( Q P' - 2 P Q') -  3Q^2}{12P}
- R, \Eq{13}\ee
where the right hand side is evaluated at  $z = \varphi ^{-1} (x)$.  In the
physical  coordinate, the associated algebraic wave functions will then take the form
\be
\psi (x) = \mu \left(\varphi ^{-1} (x)\right) \cdot\chi \left(\varphi ^{-1} (x)\right),
\Eq{32}\ee where
$\chi (z)$ is a polynomial of degree at most $n$.

The canonical form \eq1 of a quasi-exactly solvable differential operator
is not unique,  since there is a ``residual'' symmetry group which preserves
the Lie algebra $\fg_n$.  Not  surprisingly, this group is $\GLR2$, which acts
on the (projective) line by linear fractional  (M\"obius) transformations
\be z \longmapsto w = {\alpha z + \beta \over \gamma z+\delta },\qquad A =
\pmatrix{\alpha  &\beta
\cr\gamma &\delta \cr},\quad \det A = \Delta = \alpha \delta -\beta \gamma
\ne  0.\Eq8\ee To describe the induced action of the transformations \eq8 on
the quasi-exactly solvable  operators
\eq1, we first recall the basic construction of the finite-dim\-ension\-al
irreducible  rational representations of the general linear group
$\GL{2,\allowbreak\R}$.

\Df7  Let $n\geq 0$, $i$ be integers.  The irreducible multiplier
representation $\rho_{n,i}$  of $\GLR2$ is defined on the space $\CP_n$ of
polynomials of degree at most 
$n$ by the transformation rule
\be P(w)\mapsto\Phat(z) = (\alpha \delta -\beta \gamma )^i\,(\gamma z+\delta )^n \,P\left ({\alpha z
+\beta \over \gamma z+\delta }\right ),\qquad P \in  \CP_n.\Eq7\ee

The multiplier representation
$\rho _{n,i}$ has infinitesimal generators given by the  differential operators \eq3
combined with the operator of multiplication by $n+2i$  representing the diagonal subalgebra (center)
of $\glR2$.  The  action \eq8 induces an  automorphism of the Lie algebra $\fg_n$, which is
isomorphic to the representation $\rho  _{2,-1}$, and, consequently, preserves the class of
quasi-exactly solvable operators  associated with the algebra $\fg_n$.  Moreover, the
corresponding gauge action
\be\widehat T(z) =  (\gamma z + \delta )^n \cdot T(w) \cdot (\gamma z + \delta )^{-n},\qquad
w = {\alpha z + \beta \over \gamma z+\delta }\Eq9\ee
will preserve the space of quasi-exactly solvable operators \eq1.  Identifying the  operator
$T$ with the corresponding quartic, quadratic and constant polynomials $P$, $Q$, $R$,  we
find that the action \eq9 of $\GLR2$ on the space of quasi-exactly solvable second-order
operators is  isomorphic to the sum of three irreducible representations, $\rho _{4,-2}
\oplus \rho _{2,-1} \oplus
\rho _{0,0}$; the quartic $P(z)$ transforms according to $\rho _{4,-2}$, the  quadratic $Q(z)$
according to $\rho _{2,-1}$, while $R$ is constant.  Finally, the  associated module, which is just
the space of polynomials $\CP_n$, transforms  according to the representation $\rho_{n,0}$.

Using the
action of $\GLR2$, we can place the gauged operator \eq1 into a simpler  canonical form, based
on the invariant-theoretic classification of canonical forms for real  quartic polynomials,
\cite{Gurevich}, \cite{GKOnorm}.
\Th5 Under the representation $\rho _{4,-2}$ of $\GLR2$, every nonzero real quartic  polynomial
$P(z)$ is equivalent to one of the following canonical forms:
\ba
&\nu (1-z^2)(1-k^2z^2),\qquad \nu (1-z^2)(1-k^2 + k^2z^2),\qquad \nu (1+z^2)
(1+k^2  z^2),&
\nonumber\\ &\nu ( z^2 -   1),\quad \nu ( z^2 +   1),\quad  \nu  z^2,\quad \nu
(z^2+1)^2,\quad z,\quad  1,&\Eq{1000}\ea
where $\nu $ and $0 <  k < 1$ are real constants.

The solution to the normalizability problem, which the interested reader can find in
\cite{GKOnorm}, begins with a detailed analysis of
the elliptic  integral \eq{10}. It is found, first of all, that the class of quasi-exactly 
solvable potentials naturally splits into two subclasses --- the periodic
potentials, which are  never normalizable, and the non-periodic potentials,
which are sometimes normalizable. Tedious but direct calculations based on
\eq{10}, \eq{11}, \eq{13}, and \eq{32}, produce then the explicit change of variables,
the potential, and the form of the algebraic eigenfunctions for the above normal forms in 
physical coordinates.  Each of the classes of potentials is a linear combination of four 
elementary and/or elliptic functions, plus a constant which we absorb into the
eigenvalue.   The potentials naturally divide into two classes, which are listed in the
following two  Tables. In each case, the four coefficients are not arbitrary, but satisfy
a single complicated  algebraic equation and one or more inequalities; see \cite{GKOnorm}
for the details. 

First, the periodic quasi-exactly solvable potentials correspond to the cases when the real roots 
of $P$ (if any) are simple, of which there are five cases in \eq{1000}.
The explicit formulas for the corresponding potentials  follow; they depend on two real
parameters $\alpha>0$ and $k\in(0,1)$. In the first three cases,
the corresponding potentials are written in terms of the  standard Jacobi elliptic
functions of modulus $k$,
\cite[vol.~2]{Bateman}.  Also, as  remarked above, the coefficients $A,B,C,D$ are not arbitrary,
although the explicit constraints are too complicated to write here.
\vskip 20pt minus 10pt
\ctitle  Periodic Quasi-Exactly Solvable Potentials.
$$\ntable{\dn^{-2} \alpha \,x \bigl(A\sn \alpha \,x+ B\bigr)  +  \cn^{-2}  \alpha \,x 
\bigl(C\sn \alpha \,x+ D\bigr)\\
\dn^{-2} \alpha \,x \bigl(A\cn \alpha \,x+ B\bigr)  + \sn^{-2}  \alpha \,x \bigl(C\cn 
\alpha \,x+ D\bigr)\\ A \cn \alpha \,x \sn \alpha \,x+ B  \cn^2 \alpha \,x  +C
\dn^{-2}  \alpha \,x \bigl( 
\cn \alpha \,x \sn \alpha \,x + D \cn^2 \alpha \,x \bigr)\\ A \sin^2 \alpha \,x + B
\sin \alpha \,x +C \tan \alpha \,x \sec \alpha \,x + D \sec^2 
\alpha \,x\\ A \cos 4 \alpha \,x + B \cos 2\alpha \,x +C \sin 2\alpha \,x + D \sin 4
\alpha \,x}$$
\vskip 10pt

Note that the  potentials in cases 1, 2
and 4 have singularities unless $C = D = 0$.  In Cases 3 and 5, the  potential has no
singularities, reflecting the fact that in these cases $P(z)$ has no real roots. Case 3
includes the Lam\'e equation, \cite[vol.~3]{Bateman}, and Case 4 with $A=B=0$ is the trigonometric
Scarf potential.

The non-periodic potentials correspond to the cases with one or two multiple roots. The
explicit formulas for the corresponding potentials follow (again, $\alpha >0$ is a real
constant).
\vskip 20pt minus 10pt
\ctitle Non-periodic Quasi-Exactly Solvable Potentials.
$$\ntable{A \sinh^2 \alpha \,x + B \sinh \alpha \,x +C \tanh \alpha \,x \sech \alpha 
\,x + D \sech^2 \alpha \,x\\
 A \cosh^2 \alpha \,x + B \cosh \alpha \,x + C \coth \alpha \,x \csch \alpha \,x + D 
\csch ^ 2 \alpha \,x\\ A e^{2\alpha \,x} + B e^{\alpha \,x} + Ce^{-\alpha \,x} + D
e^{-2\alpha \,x}\\ Ax^6 + Bx^4 + C x^2 + {D\over x^2}\\ A x^4 + B x^3 + C x^2 + D x}$$
\vskip 10pt
In cases 2 and 4, the potential has a singularity at  $x=0$ unless $C+D = 0$
(Case 2) or $D = 0$ (Case 4).  The nonsingular potentials in Case 4 are the anharmonic
oscillator  potentials discussed in detail in \cite{Turbiner}, \cite{Shifcft}.  The
algebraic constraints  satisfied by the coefficients are given in \cite{GKOnorm}.

According to Turbiner, \cite{Turbiner}, a potential is {\it exactly solvable} if it does
not  explicitly depend on the discrete ``spin'' parameter $n$, since, in this case, one can find 
representation spaces of arbitrarily large dimension and thereby (if the algebraic
eigenfunctions are normalizable) produce  infinitely many eigenvalues by algebraic methods.  Note
that since the gauge transformation 
\eq{10}, \eq{11}, can explicitly depend on $n$, exact solvability cannot be  detected in
the canonical coordinates, but depends on the final physical form of the  operator.  The exactly
solvable nonperiodic potentials are characterized by the condition 
$A=B=0$, and, in case 3, $C=D=0$.  In Case 2, there is an additional
inequality, $|C| \leq D  +
\f4\alpha^2 $, to be satisfied.  The exactly solvable potentials include the (restricted)
P\"oschl--Teller and Scarf potentials (Case 1), the Rosen--Morse II potential (Case 2), the Morse
potential (Case 3), the radial harmonic oscillator  (provided $D = l(l+1)$, $l\in \N$) (Case 4),
and the  harmonic oscillator (Case 5).

Analysis of the explicit formulas for the eigenfunctions based on \eq{32} yields a 
complete set of conditions for the normalizability of the non-periodic potentials, which
can be written explicitly in terms of the coefficients of the quadratic polynomial
$Q(z)$. It is then possible to deduce explicit, general normalizability
conditions on the Lie-algebraic  coefficients $c_{ab}$ and $c_a$ by using the
fact that such conditions must be invariant under the  action of the group
$\GLR2$.  Therefore, normalizability conditions can be written in terms  of the
classical joint invariants and covariants of the pair of polynomials $P$,
$Q$. See \cite{GKOnorm} for the details and a complete list of invariant
normalizability conditions.

\Section{two}Two-Dimensional Problems.
There are a number of additional difficulties in the two-dimensional problem
which do not appear in  the scalar case.  First, there are several different
classes of quasi-exactly solvable Lie algebras  available.  Even more important
is the fact that, according to \th{edo.2}, there are nontrivial {\it closure 
conditions} which must be satisfied in order that the magnetic one-form
associated with a given second-order differential operator be closed, and hence
the operator be equivalent under a gauge transformation 
\eq{qes.2} to a Schr\"odinger operator.  Unfortunately, in all but trivial
cases, the closure conditions  associated with a quasi-exactly solvable
operator \eq{qes.9} corresponding to the generators of one of the
quasi-exactly solvable Lie algebras on our list are {\it nonlinear algebraic
equations} in the  coefficients $c_{ab}$, $c_a$, $c_0$, and it appears to be
impossible to determine their general  solution. Nevertheless, there are useful
simplifications of the general closure conditions which can  be effectively used
to generate large classes of planar quasi-exactly solvable and exactly solvable 
Schr\"odinger operators, both for flat space as well as curved metrics.

Suppose that the Lie algebra $\fg$ is spanned by linearly independent
first-order differential operators as in \eq{qes.20}. The
{\it closure conditions} $dA = 0$ for a Lie-algebraic second-order differential operator
\eq{qes.9} are then equivalent to the solvability of the system of partial
differential equations
\ba  \sum_{a,b=1}^r c_{ab} \xi^{ai} \sum_{j=1}^n \left (\xi^{bj}\pd \tau  {x^j}
+ \pd{\xi  ^{bj}}{x^j}\right ) = \sum_{a=1}^r \xi^{ai} \left [2\sum_{b=1}^r
c_{ab} \eta^b +c_a 
\right],\nonumber\\\rg id,&&\Eq{22}\ea  for a scalar function $\tau (x)$,
given by $ \tau = 2\sigma+\f2\log\det (g_{ij})$ in terms of the gauge factor 
$e^\sigma $ required to place the operator in Schr\"odinger form. The closure
conditions \eq{22} are  extremely complicated to solve in full generality, but a
useful subclass of solutions can be obtained  from the {\it simplified closure
conditions}
\be
\sum_{i=1}^n\left( \xi^{ai}\pd \tau {x^i} + \pd{\xi^{ai}}{x^i} \right) - 2\eta^a = k^a,\qquad
\rg  ar. \Eq{23}
\ee 
where $k^1,\ldots ,k^r$ are constants.  Any solution $\tau(x)$ of equations
\eq{23} will generate an  infinity of solutions to the full closure conditions
\eq{22}, with $c_{ab}$ arbitrary, and
$c_a = \sum_b c_{ab}k^b$.  The case $k^a = 0$ and $\fg$ semi-simple was
investigated in 
\cite{MPRST}.  Although the simplified closure conditions can be explicitly
solved for such Lie algebras, with the  exception of $\so3$, their solutions are
found to generate quasi-exactly solvable Schr\"odinger  operators that are {\it
not\/} normalizable, and hence of limited use. Note that even when the 
simplified closure conditions do not have any acceptable solutions, the {\it
full\/} closure conditions \eq{22} may be compatible and may give rise to
normalizable operators, as shown in \cite{GKOii} and \cite{GKOrqes}. Here we
shall limit ourselves to a few examples.

Consider, in the first place, the Lie algebra $\fg\simeq\sL2 \oplus \sL2$
spanned by the  first-order differential operators 
\ba  &J^1 = \px,\quad J^2 = \qy,\quad J^3 = x\px,&\nonumber\\
&J^4 = y\qy,\quad J^5 = x^2\px-nx,\quad J^6  = y^2\qy-my,&\ea
where $n,m\in \N$. The particular choice 
\ba \left(c_{ab}\right) &=&  \pmatrix{2 & 1 & 0 & 0 & 0 & 1 \cr 1 & 2 & 0 & 0 &
1 & 0 \cr 0 &  0 & 3 & 0 & 0    & 0 \cr 0 & 0 & 0 & 3 & 0 & 0 \cr 0 & 1 & 0 & 0
& 1 & 1 \cr 1 & 0 & 0 & 0 & 1 & 1},\\
\left(c_{a}\right) &=& \Bigl( 0,0,-(1 + 4n),-(1 + 4m),0,0 \Bigr),\qquad
c_0 = \fr34+ m^2  +  n^2,\ea
of Lie-algebraic coefficients lead to a quasi-exactly solvable Hamiltonian with
Riemannian metric
\be  g^{11} = ( 1 + x^2 ) (2+x^2),\quad g^{12} = ( 1 + x^2 ) ( 1 + y^2 ),\quad
g^{22} = ( 1 + y^2 )  (2+y^2),\ee
which has complicated curvature, and potential
\aeq{4V &=& -y^2 -{( 1 + 2n ) ( 3 + 2n ) \over 1 + x^2}- {( 1 + 2m ) ( 3 + 2m )
\over1 + y^2}
\nonumber\\ && - {17 + 12y^2 - {y^4} + 2xy(6 + 5y^2 ) \over 3 + x^2 +
y^2}\nonumber\\ &&\hbox{}+ {5( 3 + 2xy) ( 1 +  y^2 ) ( 2 + y^2 ) \over ( 3 + x^2
+ y^2 ) ^2}.}

Consider next the $\so{3,1}$ Lie algebra generated by the operators
\ba &J^1=\dx,\quad J^2=\dy,\quad J^3=x\dx+y\dy,\quad
J^4=y\dx-x\dy,&\nonumber\\ &J^5=(x^2-y^2)\dx+2xy\dy-2nx,\quad
J^6=2xy\dx+(y^2-x^2)\dy-2ny,&
\ea with $n\in\N$ a positive integer. Notice that this algebra, although of
course inequivalent to any of the {\it real} normal forms in the
classification of \cite{GKOrqes}, is {\it complex}-equivalent to the
$\sL2\oplus\sL2$ Lie algebra in the previous example under the complex change
of variables $z=x+i\,y$, $w=x-i\,y$.

The second-order differential operator
\ba 
\lefteqn{-T = 
\alpha\left[\JJ1+\JJ2\right]+\beta\JJ3+\gamma\JJ4}\nn\\
&&\qquad\mbox{}+\lambda\left[\JJ5+\JJ6\right]+4\beta(1+2n)^2-4\gamma
\ea
satisfies the closure conditions, and is therefore equivalent to a
Schr\"odinger operator on the manifold with contravariant metric tensor
$(g^{ij})$ given by
\ba g^{11} &=&\alpha  + \beta \,x^2 + \gamma \,y^2 + 
              \lambda \,( x^2 + y^2)^2,\nonumber\\ g^{12} &=& (\beta  -
\gamma ) \,x\,y,\nonumber\\ g^{22} &=&\alpha  + \gamma \,x^2 + \beta
\,y^2 + \lambda \,(x^2 + y^2)^2.
\ea The Gaussian curvature is
\be
\kappa = {(-\beta  + 3\gamma  ) (\alpha ^2 + \lambda ^2 r^8) +  
    2( \beta \gamma  + 4\alpha \lambda  ) r^2 (\alpha +\lambda r^4)
  +  2\alpha  \lambda( 5\beta  + \gamma  ) r^4
\over ( \alpha  + \gamma \,r^2 + \lambda \,r^4 )^{2}},
\ee with $r^2 = x^2+y^2$.   If the parameters $\alpha$, $\beta$, $\gamma$ and
$\lambda$ are positive, then the metric is non-singular and positive definite
for $(x,y)$ ranging over all of $\R^2$. The fact that the closure conditions
are satisfied guarantees the existence of a gauge factor $\mu$ such that
$H=\mu\,T\,\mu^{-1}$ is a Schr\"odinger operator; see \cite{GKOrqes} for an
explicit formula for the gauge factor, which is too complicated to be given
here. The expression for the potential $V$ is
\ba 4 V&=&
\frac{16\,\alpha \,\beta \,n\,( 1 + n )  + 
      r^2\,\left[ \beta ^2\,( 3 + 16\,n + 16\,n^2) 
      -4\,\alpha \,\lambda \,( 3 + 8\,n + 4\,n^2 )\right]}{
         \alpha  + \beta \,r^2+ \lambda \,r^4}\nonumber\\
  &&\qquad \mbox{}+\frac{5\,( \beta  - \gamma ) \,
       ( 4\,\alpha \,\lambda - \gamma ^2)  + 
      3\,\lambda \,( 2\,\beta \,\gamma  - 3\,\gamma ^2+ 
         4\,\alpha \,\lambda  ) \,r^2}{
    \lambda \,( \alpha  + \gamma \,r^2 + \lambda \,r^4)}\nonumber\\
  &&\qquad {}-\frac{5\,( \beta  - \gamma) \,
      (4\,\alpha \,\lambda  -\gamma^2) \,
      ( \alpha  + \gamma \,r^2) }{ 
    \lambda \,( \alpha  + \gamma \,r^2 + \lambda \,r^4)^2}.
\ea Since the potential is a function of $r$ only, it is natural to look for
eigenfunctions of
$H$ which depend on $r$ only. When this is done, it can be shown that one ends
up with  an effective Hamiltonian on the line which is an element of the
enveloping algebra of the standard realization of $\sL{2,\R}$ in one
dimension. Thus, no new \qes. one-dimensional potentials are obtained by
reduction of the above \qes. $\so{3,1}$ potential.  This lends additional
support to the observation that reduction of two-dimensional  quasi-exactly
solvable Schr\"odinger operators does not lead to any new one-dimensional
quasi-exactly solvable operators.

Non semi-simple Lie algebras can also yield interesting examples of
quasi-exactly solvable potentials. Indeed, let $\fg$ be the Lie algebra
spanned by the first-order differential operators
\be  J^1 = \px,\quad J^2 = \qy,\quad J^3 = x\px,\quad J^4 = x\qy,\quad J^5 =
y\qy, \Eq{50}
\ee
and
\be J^6 = x^2\px+ (r-1)xy\qy-nx, \quad J^{6+i} = x^{i+1}\qy, \qquad\rg
i{r-2}.
\ee
The module  $\CN$ is now spanned by the monomials $x^iy^j$ with
$i+(r-1)j\le n$. For
$m\in \N$, $\alpha, \beta > 0$, the  Schr\"odinger operator with metric
\be  g^{11} = \alpha x^2+\beta ,\quad g^{12} = (1+m)\alpha xy,\quad 
g^{22} = (\alpha x^2+\beta )^m + \alpha (1+m)^2y^2,\ee 
and potential
\be V = -\;{\lambda \alpha \beta (1+m)^2 (\alpha x^2+\beta )^m \over (\alpha x^2+\beta )^{1 + m} + \alpha \beta (1+m)^2 y^2
}, \qquad  m\le r-1\ne 2(m+1), \ee
is normalizable and quasi-exactly solvable
with respect to $\fg$, provided that the parameter 
$\lambda$ is large enough. The metric in this case has constant negative
Gaussian curvature $\kappa  = -\alpha $. Furthermore, since the potential $V$ does not
depend on the cohomology parameter $n$, the  above Hamiltonian is {\it exactly
solvable}. Moreover, the potential is also independent of $r$; hence  we have
constructed a single exactly solvable Hamiltonian which is associated to an
infinite number  of inequivalent  Lie algebras of arbitrarily large dimension.

\Section{matrix}Matrix Schr\"odinger operators.
We shall outline in this Section how to extend the notion of quasi-exact
solvability to matrix Hamiltonians in one dimension. By definition, a {\it
matrix Schr\"odinger operator} (or {\it matrix Hamiltonian}) is a $N\times N$
matrix of second-order differential operators of the form
$\CH = -\dx^2+V(x)$, where $V(x)$ is a $N\times N$ Hermitian matrix of
functions. If
we drop the restriction that $V$ be Hermitian,
then $\CH$ will be called a {\it Schr\"odinger-like}\/ operator. Here we
shall be mainly interested in spin
$1/2$ matrix Hamiltonians, that is in the case $N=2$; see \cite{FGR} for
the case of arbitrary $N$.

The notion of equivalence we shall use for matrix differential operators is
the same as in the scalar case (\cf \df{qes.1}), where the gauge factor
$e^{\sigma(x)}$ is replaced now by an arbitrary invertible matrix $U(x)$.
Suppose, as in the previous sections, that a Schr\"odinger operator
\be\CH(x)=U(z)\cdot  T(z)\cdot U(z)^{-1},\qquad x=\varphi(z),\ee
is equivalent to a matrix differential operator
$T(z)$ preserving a finite-dimensional subspace
$\CN\subset\C^\infty(\R)\oplus\C^\infty(\R)$ of smooth two-component wave
functions, \ie $T(\CN)\subset\CN$. It
follows that $\CH$ will restrict to a linear operator in the
finite-dimensional space $U\cdot\CN|_{z=\varphi^{-1}(x)}$, and therefore
$\dim\CN$ eigenfunctions of $\CH$ will be computable by purely
algebraic methods. We shall say that $\CH$ is {\it quasi-exactly solvable;} as
in the scalar case, the algebraic eigenfunctions of $\CH$ should obey
appropriate boundary conditions, like for instance square integrability.

Thus, to find examples of quasi-exactly solvable matrix Hamiltonians we have to
solve the following two problems:

\begin{enumerate}
\item Characterize all second-order matrix differential operators leaving
invariant a finite-dimensional space of smooth functions
\item Find when a second-order matrix differential operator satisfying the
previous condition is equivalent to a matrix Schr\"odinger operator
\end{enumerate}

The first of these problems is clearly too general. In the scalar case, it is
replaced by the simpler (but still highly nontrivial) problem of constructing
second-order differential operators belonging to the enveloping algebra of a
Lie algebra of \qes. first order differential operators, which in turn leads to
the classification of such Lie algebras. In the matrix case this is still
too general, since Lie superalgebras and even more general algebraic
structures, whose complete classification is not available yet, come into play,
\cite{BrKo}, \cite{BGGK}, \cite{FGR}.

To simplify matters, the idea is to {\it fix}\/ a concrete (but still general
enough) finite-dimensional subspace of smooth wave functions and classify all
second-order matrix differential operators leaving this subspace invariant.
Following Turbiner, \cite{Tumat}, and Brihaye {\it et al.,}\/ \cite{BrKo},
\cite{BGGK}, we choose as our invariant subspace the space of all wave
functions with polynomial components. More precisely, let
\be
\CP_{m,n}=\CP_m\oplus\CP_n
\ee
be the vector space of wave functions
$\Psi(z)=\bigl(\psi_1(z),\psi_2(z)\bigr)^t$, with
$\psi_1(z)\in\CP_m$ and $\psi_2(z)\in\CP_n$ polynomials of degrees less than or
equal to
$m$ and $n$, respectively. We shall say that a matrix differential operator
$T$ is a {\it polynomial vector space preserving}\/ (PVSP) operator if it
preserves the vector space $\CP_{m,n}$ for some non-negative integers $m$ and
$n$. We shall also denote by $\CP^{(k)}_{m,n}$ the vector space of all PVSP
matrix differential operators of differential order less than or equal to $k$.
The problem is then to characterize all second-order PVSP operators, or in
other words to describe the space $\CP^{(2)}_{m,n}$ in as concise a way as
possible.

To this end, let $\De=n-m\ge0$, and consider the $2\times2$ differential
operators
\ba
&\Tp = \diag(\Jp{n-\De},\Jp{n}),\quad
\T0 = \diag(\J0{n-\De},\J0{n}),\nonumber\\
&\Tm = \diag(\Jm,\Jm),\quad
J=\frac{1}{2} \diag(n+\De,n);\nonumber\\
&\Qa = z^\al\,\sigma^- ,\qquad \Qab =
\qab{n}{\De}\,\sigma^+ \, ,\quad
\al=0,\dots ,\De,&\Eq2
\ea
with\footnote{Here, and in what follows, we have adopted the convention that a product with
its lower limit greater than the upper one is automatically $1$.}
\be
\qab{n}{\De}=\prod_{k=1}^{\De-\al} (z\Dz -n+\De-k) \,\Dz^\al\,
\roq{and}\sigma^+=(\sigma^-)^t=\bpm 0 & 1\\0 & 0\epm.\ee
Note that,
in contrast with the scalar case, for $\De>1$ the operators $\Qab$ have
differential order grater than one.

It can be easily checked that
the
$6+2\De$ operators in~\eq2 preserve $\p{n-\De,n}$, as does any any polynomial
in the above operators. We now introduce a $\Bbb{Z}_2$-grading in the set $\Cal
D_{2\times 2}$ of
$2\times 2$ matrix \dfo s as follows: an operator
\be T=\bpm a & b\\c & d\epm ,\ee
where $a$, $b$, $c$ and $d$ are differential operators, is even if $b=c=0$, and
odd if $a=d=0$. Therefore, the
$T$'s and $J$ are even and the $Q$'s and $\Qb$'s odd. This grading,
combined with the usual product (composition) of operators, endows
$\Cal{D}_{2\times 2}$ with an associative superalgebra structure. We can
also construct a Lie superalgebra in
$\Cal{D}_{2\times 2}$ by defining a generalized Lie product in the usual way:
\be
[A,B]_s=AB-(-1)^{\deg A \deg B}BA.
\ee
However, this product does not close within the vector space spanned by
the operators in
\eq2, except for $\De=0,1$ (see \cite{FGR} for the explicit commutation
relations). More precisely, for $\De=1$ the underlying algebraic structure
is the classical simple Lie superalgebra $\frak{osp}(2,2)$, \cite{Shifman},
\cite{Tumat}, whereas for $\De=0$ it is $\frak{h}_1\oplus\sL2$, where
$\frak{h}_1$ is the 3-dimensional Heisenberg superalgebra. As remarked
in \cite{BrKo}, in the latter case we can alternatively leave the grading
aside and replace $J=1$ by $\tilde J=\sigma_3$, ending up with the Lie algebra
$\sL2\oplus\sL2$. For all other values of $\De$, the Lie superalgebra of
$\CD_{2\times2}$ generated by the operators \eq2 is
infinite-dimensional, \cite{FGR}.

In spite of the fact that in general they don't generate a finite-dimensional
superalgebra, the operators \eq2 are the building blocks in the construction of
second-order matrix differential operators preserving $\p{m,n}$; in fact,
any such operator must be a polynomial in the operators \eq2, a fact first
stated by Turbiner, \cite{Tumat}:

\Th1
Let $\De=n-m\ge0$, $m\ge2$, and let $T$ be a second-order $2\times 2$ matrix
differential operator leaving the space $\p{m,n}$ invariant. Then
\be
T =p_2(\Te)+J\,\tilde p_2(\Te)+\sum_{\al=0}^\De
\Qab\,\pb_{2-\De}^\al(\Te)+
     \sum_{\al=0}^\De \Qa\, p_2^\al(\Te),\Eq{20}
\ee
with $p_k$, $\tilde p_k$, $p_k^\al$, and $\pb_k^\al$ polynomials of
degree less than or equal to $k$. If $\De=0$, the above formula is still valid
with
$J$ replaced by $\tilde J=\sigma_3$, while if $\De>2$ every $\pb^\al_k$
must be identically zero.

See \cite{FGR} for a proof, as well as for the generalization of this result
to the case of $N\times N$ matrix \dfo s of arbitrary order.

Turning now to the second of the two fundamental problems listed at the
begininng of this section, we have the following fundamental Theorem,
\cite{FGR}:

\Th2
Let $T$ be a PVSP operator in $\p{m,n}^{(2)}$, with $n\geq m\geq 2$.
Then $T$ is equivalent to a \sch.-like operator if and only if it is
of the form
\be
-T=p_4(z)\,\Dz^2+A_1(z)\,\Dz+A_0(z)\,,\Eq3
\ee
with $p_4(z)$ a polynomial of degree at most 4. The operator $T$ is
equivalent to a
\sch. operator
$-\dx^2+V(x)$ if and only if~\eq3 holds, and in addition we have:
\begin{enumerate}
\item The physical coordinate $x=\varphi(z)$ is given by
\be
x=\int^z\frac{ds}{\sqrt{p_4(s)}};\Eq{21}
\ee
\item There is an invertible
matrix
$\Ut(x)$ satisfying the differential equation
\be
\Ut_x=\Ut A,
\Eq4
\ee
with $A$ given by%
\footnote{%
Here and in what follows, a subscripted $x$ denotes derivation with
respect to $x$, while derivatives with respect to $z$ will be denoted with a
prime $'$.}
$A(x)=\left.\frac1{2\sqrt{p_4}}(A_1-\frac12p_4')\right|
{\vphantom{\frac12}}_{z=\varphi^{-1}(x)}$;
\item The matrix
\be
V=\Ut\Wt\Ut^{-1}\,,\Eq5
\ee
with
$
\Wt=-A_0\comp\varphi^{-1}+A^2+A_x
$, is Hermitian.
\end{enumerate}
The eigenfunctions of the Hamiltonian $H$ are of the form
$\Psi(x)=\Ut(x) \Phi\left(\varphi^{-1}(x)\right)$, where
$\Phi(z)$ is an eigenfunction of $T$ and $\Psi$ must satisfy appropriate
boundary conditions.

Using the previous theorem, it can be shown that a
matrix \sch. operator equivalent to a PVSP operator $T\in\p{n-\De,n}^{(2)}$
with $\De>1$ can be transformed by a
constant gauge transformation into a \sch. operator whose potential matrix
is triangular. Since triangular potentials correspond to
essentially uncoupled systems, we can restrict ourselves without loss of
generality to the cases $\De=0$ and $\De=1$.

The above theorems tell us, in principle, how to find examples of \qes. matrix
Hamiltonians possessing a finite number of algebraically computable wave
functions. The steps in the construction can be summarized as follows:

\begin{enumerate}
\item Construct the most general PVSP operator of the form \eq{20}.
This operator will depend on a finite number of parameters, namely the
coefficients of the arbitrary polynomials appearing in \eq{20}
\item Compute the indefinite integral \eq{21} to express the
physical coordinate $x$ in terms of the ``canonical coordinate" $z$
\item Solve the matrix linear differential equation \eq4 to compute
the gauge factor $\Ut(x)$
\item Compute the potential matrix $V(x)$ using equation \eq5
\item Choose the values of the free parameters in such a way that the matrix
$V(x)$ computed in the previous step is Hermitian, and that the phyisical
wave functions satisfy the appropriate boundary conditions
\end{enumerate}
\vskip10pt
Of the previous five steps, only the first two present practical
problems. Indeed, the first step requires the computation and inversion of an
elliptic integral, while the second one involves the solution of a linear
matrix differential equation with variable coefficients.

The first of the latter two problems can be solved, as in the scalar case, by
noting that there is still a ``residual" $\GLR2$ symmetry which can be
used to put $p_4$ into one of the canonical forms \eq{l.1000}, for which
the elliptic integral \eq{21} reduces to an elementary function. We refer the
interested reader to \cite{FGR} for the complete story. As to the second
problem (the solution of the matrix differential equation \eq4), the
strategy is to impose additional conditions on the matrix $A$ that enable
the explicit integration of \eq4, without being overly restrictive.

This can be
done as follows. First of all, let us rewrite \eq4 in the $z$ variable as
\be
U' = U\Ah,\Eq{10}
\ee
where
\ba
\Ah(z)=
\frac{A\bigl(\varphi(z)\bigr)}{\sqrt{p_4}}=\frac
1{2\,p_4}\bigl(A_1-\frac{1}{2}p_4'\bigr),\qquad
U(z) = \Ut\bigl(\varphi(z)\bigr).\Eq7
\ea
Suppose that $\Ah$ satisfies
\be
[\Ah(z),\int^z_{z_0}
\Ah(s)\,ds]=0\,,\Eq8
\ee
for some $z_0\in\Bbb R$. If this condition holds, we shall
say that we are in the {\em commuting case}. (It can be shown that \eq8
is indeed verified by the \qes. \sch. operator found by
Shifman and Turbiner, \cite{ShifTurb}.) In the commuting case, we can
readily integrate the gauge equation~\eq{10}, obtaining the following general
solution:
\be
U(z)=U_0\,\exp\int^z_{z_0}\Ah(s)\,ds,\roqq{where}
U_0\in\GL{2,\Bbb C}.
\ee
It can be easily shown that the most general $2\times2$ matrix $M(z)$
satisfying \eq8 must be of the form
\be
M(z) = f(z)\,M_0+g(z),
\ee
with $M_0$ a constant matrix and $f,g$ scalar functions. Using this simple
observation and the explicit formulas for the operators \eq2 for the cases
$\De=0,1$, one readily computes the matrix $\Ah(z)$
corresponding to each of the canonical forms of $p_4(z)$. It turns out that
the first three canonical forms \eq{l.1000} of $p_4$ lead to uncoupled
Hamiltonians, and therefore can be discarded without loss of generality; see
\cite{FGR} for the explicit formulas for the remaining canonical forms.

In the {\em non}-commuting case (that
is, when $[\Ah,\int^z \Ah]\neq0$), we may still be able to
integrate the gauge equation~\eq{10} explicitly by imposing other
constraints on
$\Ah$, as \eg assuming it is uncoupled. Unfortunately, however, no interesting examples
of \qes. Hamiltonians have been found so far in the non-commuting case.

Once a
solution of the gauge equation~\eq{10} has been found, the potential
matrix $V(x)$ can be immediately computed using \eq5.
There are still two conditions that we must impose: first, that the
potential matrix
$V(x)$ in \eq5 be Hermitian, and secondly, that the algebraic wave
functions satisfy suitable boundary conditions, which in the examples that
follow simply reduce to square integrability. These conditions
(particularly the Hermiticity of $V$) are still too complicated to be
solved in full generality, even in the commuting case. This is completely
analogous to what happened in the two-dimensional scalar case (\sc{two}),
where the closure conditions cannot be solved in general, and we are
therefore restricted to obtaining particular solutions. In the same
spirit, we shall exhibit now two examples of
\qes. spin $1/2$ matrix Hamiltonians, all of which belong to the commuting case
described above; see \cite{FGR} for additional examples and further details.
Interestingly, both of these examples correspond to the case $\Delta=1$;
indeed, in the commuting case with $\Delta=0$ every potential we have obtained
turned out to be either non-normalizable, singular or diagonalizable by a
constant gauge transformation.

For our first example we consider the PVSP operator
\ba
\lefteqn{-T=(\T0)^2+2 a_2\Tp+2(n+1)\T0-2\,J\T0+2
b_1\,\Qb_0}\nn\\ &&\quad\mbox{}-2 b_1\,Q_0\T0
-(3n+1)b_1\,Q_0-4 a_2 b_1 Q_1-(2\hat
d_0+n+\frac 12)J\,,
\ea
with all the parameters real. In this example $p_4=z^2$, so that we
are in case 6 of \th{l.5}. Solving
equation~\eq{21} for $z$ we immediately obtain
$z=e^x$. The gauge factor reads:
\be
U(z)=\sqrt z\,e^{a_2 z}\bpm \cos u & \sin u \\ -\sin u & \cos u
\epm\,,
\roqq{where} u=b_1\log z\,.
\ee
The potential is given by
\ba v_j&=&-(2n+1)a_2\,e^x+a_2^2\,e^{2x}
+(-1)^j\bigl(\al(x)\cos(2b_1x)-\beta(x)\sin(2b_1x)\bigr),
\nn\\
v&=&\al(x)\sin (2b_1x)+\beta(x)\cos (2b_1x)\,,\vspace{.2cm}
\ea 
where $j=1,2$, and
\be
\al(x)=-\frac{\hat d_0}2+a_2\,e^x\,,\qquad
\beta(x)=(2n+1)\,b_1+2a_2b_1\,e^x\,.
\ee 
It is easily verified that the expected value of the
potential is bounded from below, \ie
\be
\langle\Psi,V\Psi\rangle\geq c\,\|\Psi\|^2\,,
\qquad\ro{with}\quad\Psi\in L^2(\Bbb R)\oplus L^2(\Bbb R)\,,\Eq{12}
\ee
for some $c\in\Bbb R$.
(Note, however, that even in this case the amplitude of the
oscillations of $v(x)$ tends to infinity as
$x\rightarrow+\infty$.) Finally, the condition $a_2<0$ is necessary
and sufficient to ensure the square integrability of the
eigenfunctions $\Psi(x)$.

As our last example, we consider:
\ba 
\lefteqn{-T=\Tm\T0+2 a_1\T0+(2 a_0+n-\frac
12)\,\Tm-J\Tm+2 b_1\,\Qb_0-2 b_1\,Q_0\T0}\nn\\ &&\quad
\mbox{}-b_1\,(4
a_0+3n+1)\,Q_0-4 a_1 b_1 Q_1+2(2\hat a_0-a_1)J\,,
\ea
where all the coefficients are real, and $b_1\neq 0$.
Since $p_4=z$ (case 7), we have $z=x^2/4$. The gauge factor is
chosen as follows:
$$ U(z)=z^{a_0}\,e^{a_1 z}
\bpm \cos b_1 z & \sin b_1 z \\ -\sin b_1 z & \cos b_1 z \epm\,.
$$ 
The entries of the potential $V(x)$ are given by
\ba 
v_j&=&\frac{2a_0(2a_0-1)}{x^2}+\frac
14\,(a_1^2-b_1^2)\,x^2
      +(-1)^j\big(\hat a_0\,\cos\frac{b_1
x^2}2\,-\,\al(x)\,\sin\frac{b_1 x^2}2\big)
\,,\nn\\\vspace{.2cm}
v&=&\hat a_0\,\sin\frac{b_1
x^2}2\,+\,\al(x)\,\cos\frac{b_1 x^2}2\,,\vspace{.2cm}
\ea
with $j=1,2$, and $\al(x)$ is defined by
$$
\al(x)=\frac{b_1}2\,(4a_0+4n+1+a_1 x^2)\,.
$$ 
We first note that the potential is singular at the origin
unless $a_0=0,\,1/2$.
Let us introduce the parameter $\la=2a_0-1$, in
terms of which the coefficient of $x^{-2}$ in $v_j$ is $\la(\la+1)$. If
$\la$ is a non-negative integer $l$, we may regard
\be
(-\Dx^2+V(x)-E\,)\PSI(x)=0\,,\qquad
0<x<\infty\,,\Eq{13}
\ee
as the radial equation obtained after separating
variables in the three-di\-men\-sion\-al Schr\"od\-inger equation with a
spherically symmetric Hamiltonian given by
$$
\hat H=-\De+U(r)\,,\roqq{with}
U(x)=V(x)-\frac{l(l+1)}{x^2}\,,
$$ 
where $\De$ denotes the usual flat Laplacian. Given a
non-negative integer $l$ and a spherical harmonic
$Y_{lm}(\theta,\phi)$, if $\PSI$ is an eigenfunction for the
equation~\eq{13} satisfying
\begin{equation}\Eq{14}
\lim_{x\rightarrow 0^+}\PSI(x)=0\,,
\end{equation} then
$$
\hat\Psi(r,\theta,\phi)=\frac{\PSI(r)}r\,Y_{lm}(\theta,\phi)
$$ 
will be an eigenfunction for $\hat H$ with angular momentum $l$.
If $\la$ is not a non-negative integer, we shall
consider~\eq{13} as the radial equation for the singular
potential $U(r)=V(r)$ at zero angular momentum. The potential
$U(r)$ is physically meaningful, in the sense that the Hamiltonian
$\hat H$ admits self-adjoint extensions and its spectrum is bounded
from below, whenever $\la\neq -1/2$, \cite{GaPa90},
\cite{GKOnorm}. The boundary condition~\eq{14} must be
satisfied in the singular case for {\em all} values of $\la$. This
boundary condition is verified if and only if
$a_0>0$. The
expected value of the potential is bounded from below, that is,
equation~\eq{12} holds, if and only if
$$
\left|\frac{a_1}{b_1}\right|>1+\sqrt 2\,.
$$
Finally, the conditions
$$
a_0\geq 0\,,\qquad a_1<0\,,
$$
guarantee the square integrability of the eigenfunctions
$\PSI(x)$.

\end{document}